\documentclass{aa}  
\usepackage{graphicx}
\usepackage{txfonts}
\def\teff{$\rm T_{\rm eff}$}

\def\vt{${\rm v_{\rm t}}$}

\begin{document} 

\title{The metallicity distribution in the core of the Sagittarus dwarf spheroidal: minimising the metallicity biases
           \thanks{Based on observations collected at the ESO-VLT under programs 105.20AH.001}
          }

   \author{Alice Minelli \inst{1}
          \and
          Michele Bellazzini \inst{2}
          \and 
          Alessio Mucciarelli \inst{1,2}
          \and
          Piercarlo Bonifacio \inst{3}
          \and
          Rodrigo Ibata  \inst{4}
          \and
          Donatella Romano \inst{2}
          \and
          Lorenzo Monaco \inst{5}
          \and
          Elisabetta Caffau \inst{3}
          \and
          Emanuele Dalessandro \inst{2}
          \and
          Raffaele Pascale \inst{2}
          }

   \institute{
   Dipartimento di Fisica e Astronomia  “Augusto Righi", Alma Mater Studiorum, Universit\`{a}  di Bologna, Via Gobetti 93/2, I-40129 Bologna, Italy
         \and
         INAF - Osservatorio di Astrofisica e Scienza dello Spazio di Bologna, Via Gobetti 93/3, 40129 Bologna, Italy
         \and
         GEPI, Observatoire de Paris, Universit\'{e} PSL, CNRS,  5 Place Jules Janssen, 92190 Meudon, France
         \and
         Universit\'{e} de Strasbourg, CNRS, Observatoire astronomique de Strasbourg, UMR 7550, F-67000 Strasbourg, France
         \and
         Departamento de Ciencias Fisicas, Universidad Andres Bello, Fernandez Concha 700, Las Condes, Santiago, Chile
             }

\authorrunning{Minelli et al.}
\titlerunning{The metallicity distribution in the core of the Sagittarus dwarf spheroidal}

  \abstract{
  We present metallicity and radial velocity for 450 bona-fide members of the Sagittarius dwarf spheroidal (Sgr dSph) galaxy, measured from high resolution (R$\simeq 18000$) FLAMES@VLT spectra. The targets were carefully selected (a) to sample the core of the main body of Sgr dSph while avoiding contamination from the central stellar nucleus, and (b) to prevent any bias on the metallicity distribution, by selecting targets based on their Gaia parallax and proper motions. All the targets selected in this way were confirmed as radial velocity members. 
  We used this sample to derive the first metallicity distribution of the core of the Sgr dSph virtually unaffected by metallicity biases.
  The observed distribution ranges from [Fe/H]$\simeq-2.3$ to [Fe/H]$\simeq0.0$, with a strong, symmetric and relatively narrow peak around [Fe/H]$\simeq -0.5$ and a weak, extended metal-poor tail, with only $13.8\pm1.9$\% of the stars having [Fe/H]$< -1.0$.  We confirm previous evidence of correlations between chemical and kinematical properties of stars in the core of Sgr. In our sample stars with [Fe/H]$\ge -0.6$ display a lower velocity dispersion and a higher rotation amplitude than those with [Fe/H]$< -0.6$, confirming previous suggestions of a disk/halo structure for the progenitor of the system.}

 \maketitle
%
%-------------------------------------------------------------------

\section{Introduction}
The Sagittarius dwarf spheroidal galaxy \citep[][Sgr dSph, hereafter Sgr, for brevity]{Ibata1994} is the most obvious example of the ongoing disruption of a dwarf satellite into a large galaxy, the Milky Way (MW). Nowadays, we can see the remnant of the dwarf galaxy, a large low surface brightness elongated spheroid, mostly composed by unbound stars, \citep[hereafter the main body, see, e.g.,][and references therein]{vb20,Fergus2020,Delpino21,Carlberg2022}, 
and the two arms of its tidal streams wrapping around the MW \citep[hereafter the stream,][and references therein]{LM16,Iba20stream,Antoja20stream,Ramos20stream,vasi_tango}. 
The disruption of Sgr is contributing to the build-up of the MW halo in terms of dark matter, stars, and globular clusters \citep[see, e.g.,][]{Majewski03,hux2015,Hass2019,bellaz20}. The interaction with the MW appears to have left its imprint in the structure, kinematics and star formation history of the MW disc \citep[see, e.g.,][]{Laporte2019,Ruizlara2020,carr22}.

\citet{vb20} estimates that the present-day total mass of the main body is $M\sim 4\times 10^8~M_{\sun}$, with $M_{\star}\sim 1\times 10^8~M_{\sun}$ in stars, but several lines of evidence suggest that the original progenitor of the system was significantly more massive, in the range $10^{10}-10^{11}~M_{\sun}$ \citep{Lokas10,NiedersteOstholt2012,Gibbons2017,Dierickx2017,Minelli2021a,vasi_tango}.
Given the advanced stage of disruption, the chemical abundance information that low-mass stars recorded in their atmospheres since the epoch of their birth, is a key element to reconstruct the characteristics of the progenitor of the system that we observe today.
Indeed, the presence of a metallicity gradient within the Sgr main body \citep[see][for a recent thorough analysis and references]{pristine_sgr} and along the Sgr stream \citep{bellaz_grad,Chou2007,Monaco2007,Carlin2012,deBoer2014,Gibbons2017,Yang19,Hayes2020,Ramos2021} is firmly established, suggesting a complex interplay between the metallicity and kinematics that traces the combination of initial conditions in the progenitor and of the disruption process \citep{Gibbons2017,Yang19,Johnson2020,Ramos2021}.

Disappointingly enough, in spite of the increasingly detailed view of the chemical composition along the stream as well as in the main body, a robust and unbiased determination of the metallicity distribution function (MDF) of the main body is still missing.
The main body of Sgr lies at low Galactic longitude and latitude, therefore the Colour Magnitude Diagram (CMD) from which candidates members can be selected for spectroscopic follow-up is strongly affected by contamination from foreground stars from the Bulge and the Thick Disc of the MW. The combination of high mean metallicity and distance of Sgr makes it relatively easy to pick out good candidate members from the red (metal-rich) side of the red giant branch (RGB), introducing in this way an observational bias against metal-poor stars, that is somehow affecting also the most recent and thorough studies (see, e.g., \citealt[][their Appendix C in particular,]{Hasselquist2021} and also \citealt[][and references therein]{Johnson2020}).
Furthermore, several abundance analysis studies of the system were focused on the very central region of the galaxy \citep[][M17 hereafter]{Monaco2005,Bellazzini2008,Carretta2010x,Carretta2010y,Alfaro2019,Alfaro2020,Mucciarelli2017}, that hosts a complex and composite stellar nucleus, whose stellar content is not representative of the main body of Sgr \citep[][M17]{Siegel2007,Alfaro2019}. 
\\
Here we report on the results of an experiment aimed at obtaining a well sampled MDF of the core of Sgr, not affected by the metallicity biases described above (see Sect.~\ref{spectra}, for further details), and not contaminated by the population of the nuclear region.
The main new factor allowing us to get an unbiased MDF
is a selection of candidate targets for spectroscopy primarily based on {\sl Gaia} EDR3 \citep{Brown2021} parallaxes and, especially, proper motions, as done also, e.g., by \citet{vb20}, \citet{Fergus2020}, \citet{Delpino21}, \citet{Carlberg2022} and other authors. This effectively removes the MW contaminants from the CMD, providing a clean sample of high-probability Sgr members over the whole colour/metallicity range spanned by Sgr RGBs, thus avoiding biases against metal-poor stars.

As briefly reminded above, the Sgr system is very extended, also when considering only the main body  \citep[][]{vb20,Majewski03}, and a metallicity gradient with distance from the centre of the galaxy is observed at any scale \citep[see, e.g.,][]{sdgs1,Alard01,Chou2007,Majewski2013,pristine_sgr}. This means that while we now can gain a much stronger control on metallicity biases, the MDF will unavoidably depend on the radial range sampled. Indeed, none of the existing studies can claim to have obtained a MDF that is representative of the entire system, or even of the entire main body. Our case is no exception. Our choice is to focus on the core of the main body, and, in particular, on the most central part of the core, not contaminated by the stars of the nuclear star cluster. This central, non-nuclear region is supposed to be less impacted by the ongoing disruption process \citep{Delpino21}, hence, presumably, the best approximation available of the conditions near the centre of the Sgr progenitor (but see Sect.~\ref{chemod}). Our study should be considered a first step of the mapping of the MDF over the Sgr main body without metallicity biases. In the following, whenever we define our MDF as ``unbiased'', we intend that metallicity biases should have been reduced to a minimal, presumably negligible, amount.

We secured high resolution spectra for 450 Sgr members selected in this way, from which we obtained reliable and precise radial velocity (RV) and metallicity ([Fe/H]) measures, finally obtaining the desired MDF of the core of the Sgr main body.
The paper is organised as follows: in Sect.~\ref{spectra} we present the sample and report on the determination of the stellar atmospheric parameters, in Sect.~\ref{ana} we describe the analysis leading to the measure of individual RV and [Fe/H], in Sect.~\ref{resu} we show the newly derived MDF of Sgr dSph, comparing it with those of other nearby dwarf galaxies and with the predictions of chemical evolution models. We investigate also the correlation between metallicity and kinematics in our sample. Finally, in Sect.~\ref{conclu} we summarise our conclusions.

\section{Spectroscopic dataset}
\label{spectra}

\begin{figure*}[!htbp]
\includegraphics[scale=0.38]{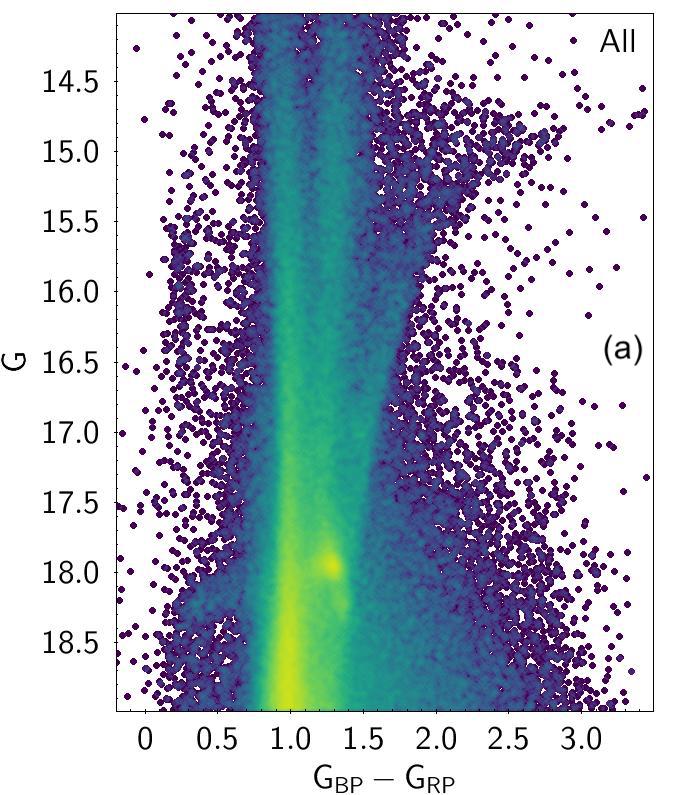}
\includegraphics[scale=0.38]{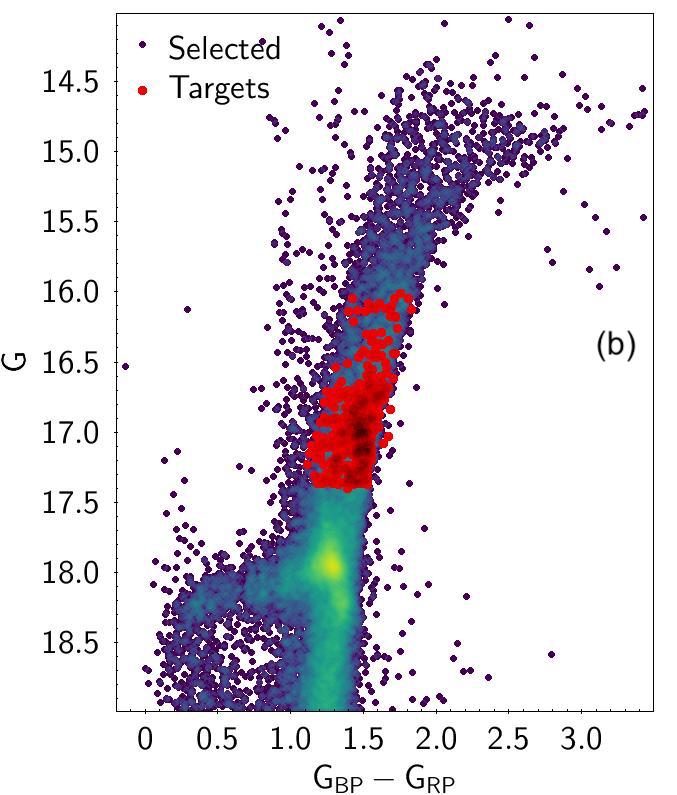}
\caption{Panel (a): Gaia DR3 CMD of a circle of radius 1.0\degr centred on the centre of Sgr dSph, with the innermost nuclear region ($R<11.5$\arcmin) excised.
The stars are colour coded according to the log of the local density.
Panel (b): the subset of the stars shown in panel (a) having proper motion within 0.5~mas/year of the systemic motion of Sgr dSph and parallax within 3$\sigma$ from 0.0~mas; in red the 450 stars that are the object of the present analysis.}
\label{cmd_sgr_flames}
\end{figure*}

The rationale of our selection is illustrated in Fig.~\ref{cmd_sgr_flames}. 
Figure~\ref{cmd_sgr_flames}a shows the CMD of a circular region within $1.0\degr$  from the centre of Sgr (coinciding with the centre of the massive globular cluster M~54) with the innermost $11.5\arcmin$ excised to minimise the contamination by the compact stellar nucleus \citep{Bellazzini2008,Carlberg2022}\footnote{According to the 2010 version of the \citet{Harris1996} catalogue the tidal radius of M~54 is $r_t=9.9\arcmin$, while \citet{Bellazzini2008} reports $r_l\simeq 10.5\arcmin$ as the limiting radius of the metal-rich component of the stellar nucleus.}. While part of the horizontal branch (at $G\simeq 18.3$ and G$_{BP}$-G$_{RP}<0.7$) and the bright asymptotic giant branch (AGB) sequence of Sgr, bending to the red from  (G,G$_{BP}$-G$_{RP}$)$\simeq(14.6,2.3)$,  are relatively clean, its red clump (G,G$_{BP}$-G$_{RP}$)$\simeq(17.9,1.3)$ and the RGB, from (G,G$_{BP}$-G$_{RP}$)$\simeq(19.0,1.0)$ to (G,G$_{BP}$-G$_{RP}$)$\simeq(14.6,2.3)$, are strongly contaminated by foreground Galactic stars. The blue side of the RGB, expected to host the most metal-poor old stars, is especially affected.

%###################################################################
%###################################################################

However, Fig.~\ref{cmd_sgr_flames}b shows how we can remove the vast majority of the contaminating stars (virtually their entirety) by selecting only stars (1) with proper motions within 0.5~mas/yr of the systemic proper motion of Sgr, as determined by \citet[][]{Helmi2018}, and (2) with parallaxes consistent with 0.0~mas within $\pm 3$ times the associated uncertainty. The first criterion selects stars with motion in the plane of the sky within $\simeq\pm 60$~km/s from the systemic motion of Sgr. This range is approximately broader than five times the typical line of sight (los) velocity dispersion in the main body \citep[][M13 hereafter]{Bellazzini2008,Majewski2013}, thus ensuring that no significant bias on the MDF can arise from kinematic cuts. The second criterion removes most of the foreground stars that happen to have proper motions within the adopted selection window, and it is justified by the fact that the parallax corresponding to the distance to the core of Sgr dSph \citep[that we assume all over the paper to be $D=26.3$, from][]{Monaco2004a} is $\omega\simeq 0.04$~mas, virtually indistinguishable from zero, within the typical uncertainty of the considered data ($err_{\omega}\simeq 0.08$ mas). From this sample is easy to select stars sampling the entire colour range spanned by the Sgr RGB, thus avoiding the biases against metal-poor stars affecting previous analyses. We also selected our targets only in the magnitude range 16.0 $<$ G $<$ 17.4 (approximately corresponding to 16.4 $<$ V $<$ 17.8), in order to avoid stars so cool to have their spectra badly affected by TiO bands \citep{Monaco2005} without introducing a colour cut, that would bias the sample against the most metal-rich bright stars. We excluded stars whose light could be possibly contaminated by close sources by selecting on the Gaia quality parameter {\tt phot\_bp\_rp\_excess\_factor} \citep{evans_dr2} according to Eq.~C.2 of \citet{linde_dr2}\footnote{ That is the following criterion on E={\tt phot\_bp\_rp\_excess\_factor}:  $1.0+0.015(G_{BP}-G_{RP})^2<E<1.3+0.06(G_{BP}-G_{RP})^2$.}. Finally, to avoid contamination of the light collected by individual FLAMES fibres from (relatively) bright sources near our spectroscopic targets, we excluded stars of magnitude G$_{\star}$ having a companion closer than $2.0\arcsec$ and brighter than G=G$_{\star}$+1.0 \citep[a common practice for this kind of observations, see, e.g.,][]{Carretta2009a}.

\begin{figure} [!b]
 \centering%
\includegraphics[width=\columnwidth]{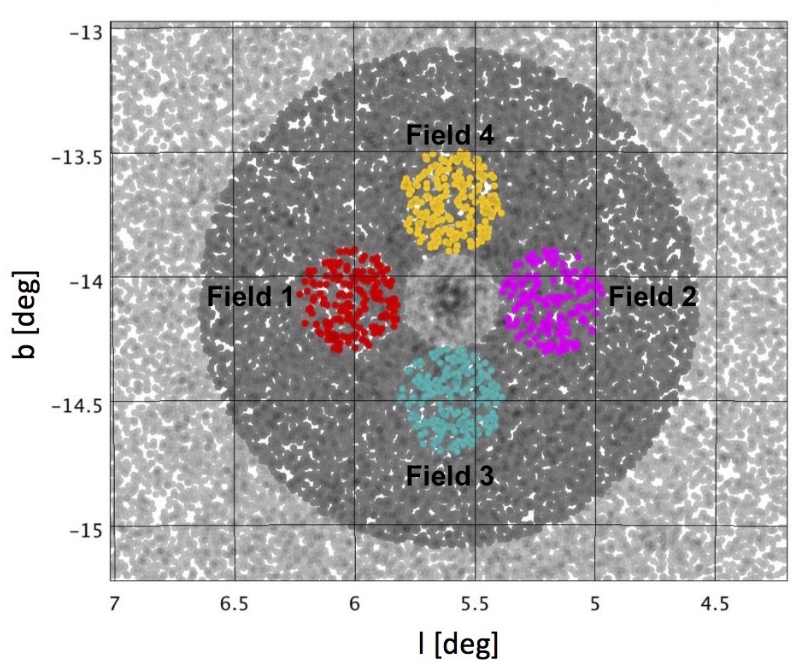}
\caption{Map of the central region of Sgr dSph. Only stars selected as likely members of the dwarf are plotted. Stars in the range $11.5\arcmin < R < 60.0\arcmin$ are plotted as dark grey filled circles. The observed spectroscopic targets in the four FLAMES fields are shown as coloured filled circles.}
\label{mappa_Sgr}
\end{figure}

From the resulting sample we extracted the stars lying in the four fields of view of the multi-object spectrograph GIRAFFE-FLAMES \citep[mounted at the Very Large Telescope of ESO,][]{Pasquini2002} located around the nuclear region of Sgr as shown in Fig.~\ref{mappa_Sgr}.
Then, the 450 stars observed in the four fields were selected by the automated fibre allocation procedure. Once again, this should prevent any bias on metallicity or radial velocity on the stars of our final sample, shown as red dots in Fig.~\ref{cmd_sgr_flames} and as dots coloured according to the specific FLAMES field in Fig.~\ref{mappa_Sgr}. 
\\
All the spectra were acquired with the GIRAFFE-FLAMES HR21 setup (spectral range 8484 -- 9001 \AA\ and resolving power $\frac{\lambda}{\Delta \lambda}\simeq 18000$). The observations were collected under the ESO program 105.20AH.001 (PI: Bellazzini), and took place between 28$^{th}$ June and 5$^{th}$ July 2021\footnote{Observations should have been carried out during the summer 2020, but they were postponed to summer 2021 because of the restrictions to ESO observatories operations due to the Covid-19 epidemic. At the epoch, the sample selection was performed on the Gaia DR2 catalogue \citep{Brown2018}. We verified that the selected stars remain bona-fide candidate members of the main body of Sgr if astrometry from Gaia EDR3 \citep{Brown2021} is used instead. Indeed, in all the paper we make use only of EDR3 astrometry and photometry \citep{Riello2021}.}. 

For each field two $t_{exp}=2775$~s exposures were acquired.
The spectra were reduced with the dedicated ESO pipeline\footnote{http://www.eso.org/sci/software/pipelines/}, that performs the bias subtraction, flat-fielding, wavelength calibration, spectral extraction and order merging. The individual exposures have been sky-subtracted using the average spectrum of some close sky regions observed at the same time of the science targets, and then they were combined in a single spectrum for each star, in order to reach a signal to noise ratio (S/N) per pixel of at least 40 for the faintest stars and 75 for the brightest ones.\\

%%%%%%%%%%%%%%%%%%%%%%%%%%%%%%%%%%%%%%%%%%%%%%%%%%%%%
%%%%%%%%%%%%%%%%%%%%%%%%%%%%%%%%%%%%%%%%%%%%%%%%%%%%%

\section{Atmospheric parameters}
\label{param}

For all the selected targets we have accurate {\sl Gaia} EDR3 photometry ($G$ and $G_{BP}$ - $G_{RP}$), from which we can obtain the atmospheric parameters in a homogeneous way.  
Effective temperatures (\teff ) have been derived using 
the (BP-RP)$_0$-\teff\ transformation by \cite{Gaiateff2021}
and adopting E(B-V) from the reddening maps of \citet{SFD98}, as re-calibrated by \citet{Schlafly2011}. The mean values of E(B-V) are 0.122 $\pm$  0.003, 0.126 $\pm$ 0.004, 0.133 $\pm$ 0.005, 0.138 $\pm$ 0.002 for fields 1 to 4, respectively (see Fig.~\ref{mappa_Sgr}). 
Surface gravities have been derived following the iterative procedure described by \citet[][their Sect.~4.1]{Lombardo2021}, 
adopting \teff\ derived above, the distance of $D=26.3$~kpc from \cite{Monaco2004a} and calculating the G-band bolometric correction BC(G) by interpolating (at fixed metallicity and \teff\ ) in a grid of theoretical BC(G) values obtained from 
ATLAS9 model atmospheres.

The microturbulent velocities \vt\ have been derived from the relation of \cite{MucciarelliBonifacio2020}, according to the log~g and the metallicities of the stars. Since the atmospheric parameters derived in this way depend on the adopted metallicity value, the procedure was repeated, updating all the parameters at each step, until convergence.
\\
Uncertainties in \teff\ are dominated by the uncertainty in the adopted colour-\teff\ transformation \citep[$\sim$80 K, see][]{Gaiateff2021}, while the contribution by photometry and reddening errors is negligible (less than 10 K).
Uncertainties in log~g are of about 0.1, including the contribution of errors in \teff, adopted distance and stellar mass.
Finally, we assume a typical error of 0.2 km/s for \vt\ according to the
uncertainties in log~g and in the adopted \vt-log~g calibration.

The final atmospheric parameters for the stars in our sample are listed in Table~\ref{info_star}, together with their coordinates, Gaia EDR3 photometry and the measured radial velocities and metallicities.

\section{Analysis}
\label{ana}
\subsection{Radial Velocity}

Heliocentric radial velocity measures were obtained with {\tt DAOSPEC} \citep{Stetson2008}, that automatically finds the centroid of spectral lines by Gaussian fitting. The final RV is the mean derived from the wavelength shift of the N measured lines and the associated uncertainty is the standard deviation divided by $\sqrt{N}$.
\\
The final spectra of 31 of the 450 observed stars were affected from particularly strong residuals of sky subtraction. For these stars we preferred to derive the RV interactively, by measuring the wavelength shift for the CaII triplet lines using the {\tt IRAF} task \textit{splot}. We conservatively assign an uncertainty of 1.0~km/s to these stars, a strong upper limit to the distribution of the measured RV uncertainties.
The uncertainties in RV  ranges from 0.2~km/s to 1.0~km/s, 80\% of the sample having ${\rm err_{RV}}< 0.5$~km/s.

%%%%%%%%%%%%%%%%%%%%% Fe analysis
\subsection{Metallicity}

The chemical abundances are then derived using our own code 
{\tt SALVADOR} that performs a $\chi ^2$ minimisation between the observed line and a grid of suitable synthetic spectra
calculated on the fly using the code {\tt SYNTHE} \citep{Kurucz2005} and varying only the abundance of the corresponding element. Model atmospheres have been calculated for each star with the code {\tt ATLAS9} \citep{Kurucz1993,Kurucz2005}. The lines are selected in order to avoid blended or saturated ones, keeping only transitions not affected by residuals of the sky subtraction. The number of Fe lines used changes with the metallicity, atmospheric parameters and noise of the spectra, going from 2 (1\% of the sample) to 19 lines (0.5\% of the sample). The median value of the number of Fe lines is 15.
\\
For the determination of the uncertainty, two main sources of error were taken into account: the error arising from the measurement procedure and that arising from the uncertainty in atmospheric parameters. The first has been computed as the standard deviation of the Fe abundance measures, divided by the square root of the number of lines used to derive the metallicity. 
The error in [Fe/H] arising from the uncertainties in the adopted parameters has been estimated by repeating the analysis of all the stars by varying the parameters of the corresponding errors, as estimated in Section \ref{param}. This uncertainty has been added in quadrature to the statistical error associated to the mean [Fe/H] of each star. The uncertainties in [Fe/H]  ranges from 0.02~dex to 0.24~dex, 80\% of the sample having ${\rm err_{[Fe/H]}}\le 0.1$~dex.

%%%%%%%%%%%%%%%%%%%%%%%%%%%%%%%%%%%%%%%%%%%%%%%%%%%%
\begin{table*}[!t]\footnotesize 
\caption{Main parameters of the target stars.}
\label{info_star}
\centering
\begin{tabular}{l c c c c c c c c c c c c}
Gaia DR3 ID&Ra&Dec&G&BP&RP&$T_{eff}$&log~g&$v_t$&RV&err&[Fe/H]&err\\
          &[deg]&[deg]&[mag]&[mag]&[mag]&\degr K&cm~s$^{-2}$&[km/s]&[km/s]&[km/s]&[dex]&[dex]\\
\hline
6760458745060096384&284.031&-30.242&16.78&17.50&15.96&4444&1.66&1.5&137.4&0.3&-0.41&0.08\\
6760462318472909312&284.057&-30.175&16.79&17.55&15.94&4341&1.60&1.5&143.1&0.3&-0.43&0.09\\
6761208779461501184&283.796&-30.150&16.82&17.58&15.98&4346&1.62&1.5&140.7&0.3&-0.44&0.10\\
6760455446525277056&283.894&-30.276&16.82&17.50&16.02&4524&1.71&1.5&150.2&0.4&-0.75&0.07\\
6760462898263096832&283.947&-30.174&16.84&17.50&16.07&4605&1.78&1.4&146.6&0.3&-0.46&0.06\\
...&&&&&&&&&&&\\
\hline 		
\hline
\end{tabular}
%\begin{tablenotes}
\tablefoot{The full version of the table is available online.}
%\end{tablenotes}
%\tablecomments{The full version of the table is available online.}
\end{table*}
%%%%%%%%%%%%%%%%%%%%%%%%%%%%%%%%%%%%%%%%%%%%%%%%%%%%

\section{Results}
\label{resu}
\subsection{Radial velocity distribution}
\begin{figure} [ht]
 \centering
\includegraphics[width=\columnwidth]{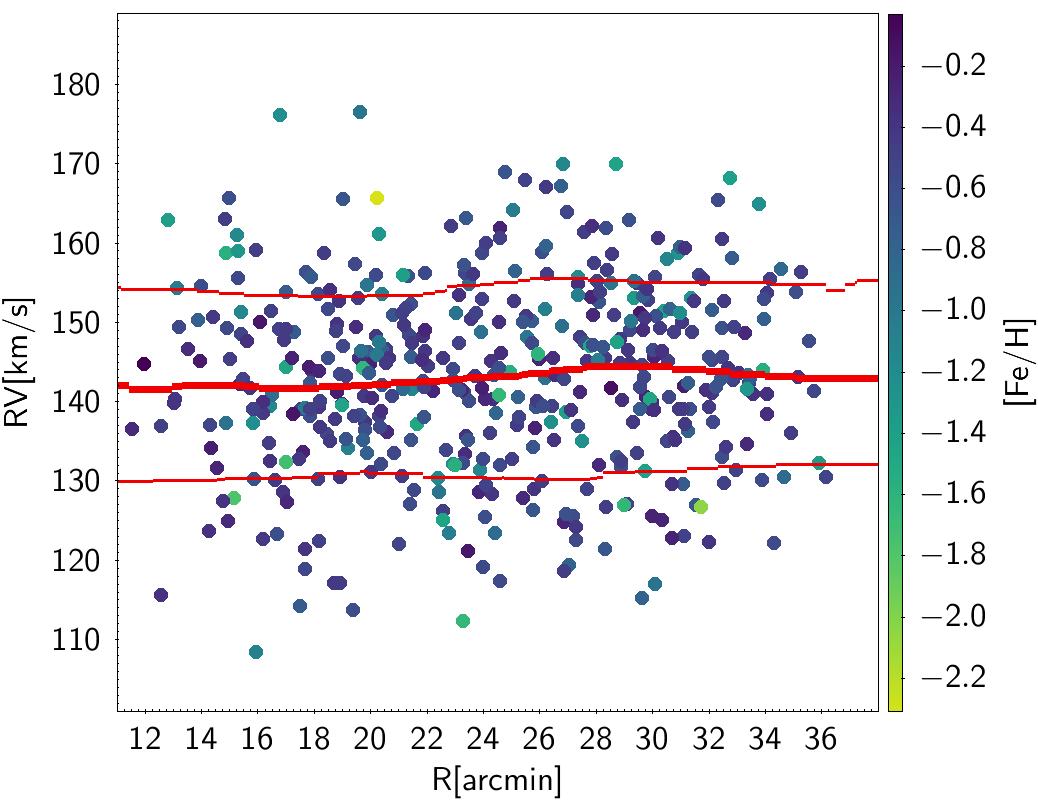}
\caption{Radial velocity of the surveyed stars as a function of the angular distance from the centre of the Sgr galaxy. Stars are colour-coded according to their iron abundance. The thick red line is the running median smoothed over 8\arcmin\ radial bins, while the thin red lines mark the 16th and 84th percentiles of the RV distribution, with the same smoothing, approximately enclosing the $\pm 1\sigma$ interval about the systemic velocity.}
\label{rv}
\end{figure}

The RV distribution of the analysed stars is shown in Fig. \ref{rv}, as a function of the angular distance from the centre of the Sgr dSph galaxy.
Following \cite{Ibata1997}, we consider bona-fide Sgr members the stars with RV between +100 km/s and +180 km/s. 
We find that all the observed stars fulfil this requirement. Therefore, the rate of success of the adopted selection of candidate Sgr member stars is 100\%, to be compared with typical rates $<$80\% for best-effort pre-Gaia purely photometric selections. For example only $\simeq$73\% of the candidate Sgr RGB stars surveyed by \citet{Bellazzini2008} were confirmed as bona-fide members of the dwarf galaxy, based on their radial velocity (843 of 1152 stars).

This result confirms that proper selections based on Gaia astrometry, in addition to preventing any metallicity bias, can be extremely efficient in picking up galaxy members, and should be adopted to maximise the scientific return of future spectroscopic surveys devoted to study Sgr stars, as, for instance, those planned with MOONS@VLT \citep{Cirasuolo2020,Gonzalez2020}. 

%%%%%%%%%%%%%%%%%%%%%%%%%%%%%%%%%%%%%%%%%%%%%%%%%%%%%
%%%%%%%%%%%%%%%%%%%%%%%%%%%%%%%%%%%%%%%%%%%%%%%%%%%%%

The radial range spanned by our data is $11.5\arcmin \la R \la 36.2\arcmin$, corresponding to $88~{\rm pc} \la R \la 277~{\rm pc}$, to be compared to the core radius of the dwarf galaxy, $r_c=224\arcmin \pm 12\arcmin$, corresponding to $r_c=1720\pm 100$~pc, as determined by \citet{Majewski03}. Hence, while avoiding the nuclear region, we are sampling the very central part of the Sgr core.

The RV distribution of Fig.~\ref{rv} is fully compatible with previous results in the literature \citep[][M13]{Ibata1997,Bellazzini2008}. The velocity dispersion profile is flat, within the uncertainties. The mean velocity and intrinsic velocity dispersion for the entire sample, estimated with the simple maximum likelihood procedure described in \citet{pm93} and \citet{walker06}, are $\langle RV\rangle= 142.9\pm0.5$~km/s and $\sigma_{int}=11.6\pm 0.4$~km/s (see also Table~\ref{tab:mcmc} for fully compatible estimates obtained with a slightly different technique).

\subsection{Metallicity Distribution}
\label{mdf}

\begin{figure} [ht]
 \centering%
\includegraphics[width=\columnwidth]{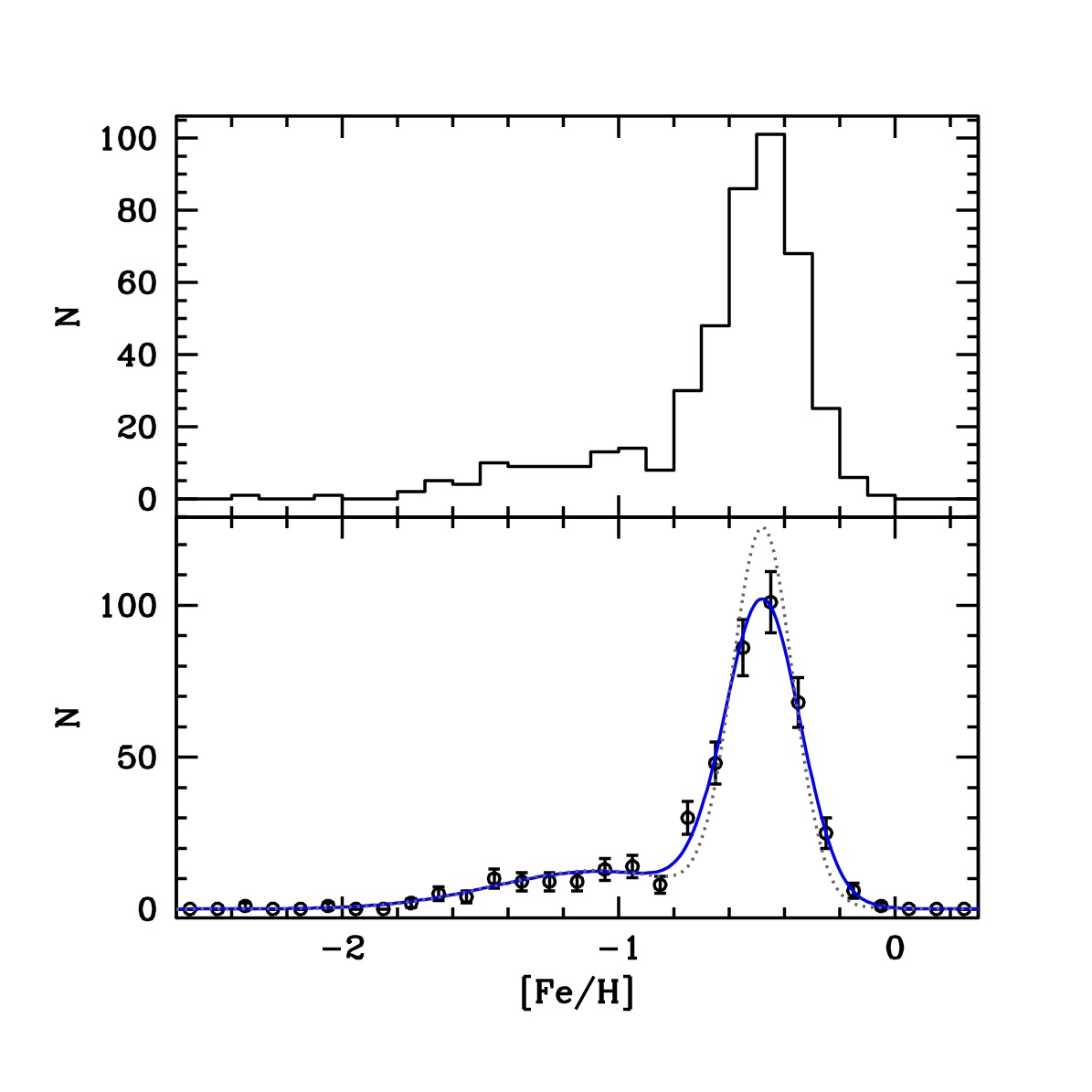}
\caption{Upper panel: metallicity distribution of the target stars displayed as a normal histogram. Lower panel: the same distribution represented as points with Poisson error bars. The grey, dotted curve is the best-fit two Gaussian model with the estimated intrinsic $\sigma$ values, while the blue continuous curve is the same model convolved with the mean uncertainty of individual [Fe/H] measures ($0.08$~dex).  }
\label{dist_met}
\end{figure}

The first unbiased MDF of the core of Sgr dSph is shown in Fig.~\ref{dist_met}. In overall agreement with previous results in the literature \citep[see, e.g.][M17, and references therein]{Bellazzini2008,Hasselquist2017,Hayes2020,Hasselquist2021}, the distribution is dominated by a strong, relatively narrow and symmetric peak at [Fe/H]$\simeq-0.5$, and displays a weak but extended tail reaching
[Fe/H]$\la-2.0$. 
\\
We used the {\tt Mclust} package \citep{mclust} within the {\tt R} environment\footnote{\tt https://www.r-project.org} to parameterise the distribution with a simple Gaussian mixture model. {\tt Mclust} uses the Bayesian Information Criterion (BIC) to select the number of mixing components. It turns out that the preferred model is made by two Gaussian components with ($\mu_1$,$\sigma_1$)=(-0.47,0.13) and ($\mu_2$,$\sigma_2$)=(-1.05,0.39), respectively, with the main metal-rich component accounting for 74.4\% of the sample and the broad metal-poor one for the remaining 25.6\%. Unfortunately {\tt Mclust} does not take into account uncertainties in individual measures, hence we recur to a maximum likelihood procedure to estimate the intrinsic values of the model parameters and the associated uncertainties. Since the [Fe/H] errors are small, the values are very similar to those derived with {\tt Mclust}, i.e., ($\mu_1$,$\sigma_1^{int}$)=($-0.478\pm 0.008$, $0.109\pm 0.007$) and ($\mu_2$,$\sigma_2$)=($-1.077\pm0.057 $, $0.368\pm 0.032$), with the fraction of stars in the first component $f_1=0.755\pm0.030$.

%%%%%%%%%%%%%%%%%%%%%%%%%%%%%%%%%%%%%%%%%%%%%%%%%%%%%%%%
%%%%%%%%%%%%%%%%%%%%%%%%%%%%%%%%%%%%%%%%%%%%%%%%%%%%%%%%

\begin{figure} [ht]
 \centering%
\includegraphics[width=\columnwidth]{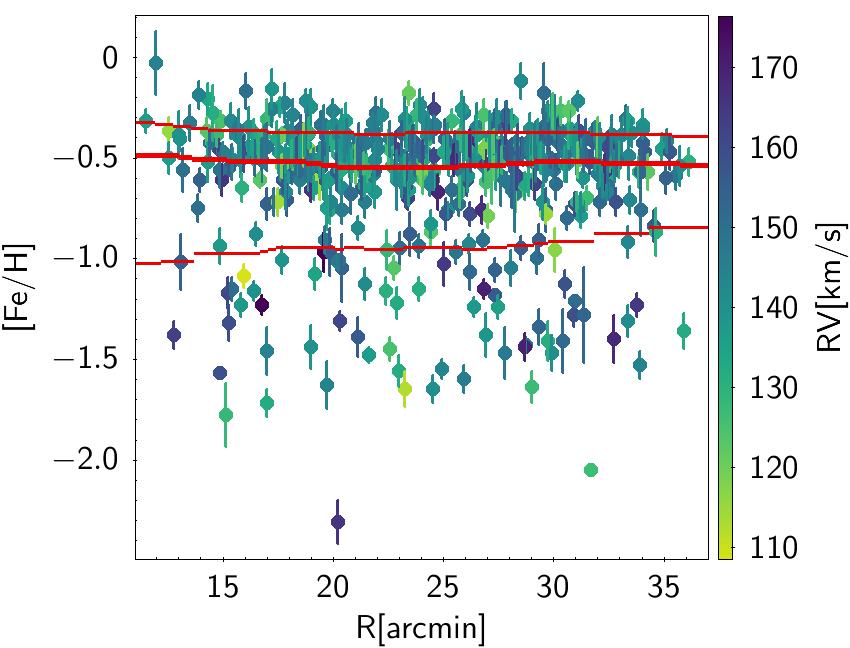}
\caption{Iron abundance of the surveyed stars as a function of the angular distance from the centre of the Sgr galaxy. Stars are colour-coded according to their RV. The thick red line is the running median smoothed over 8\arcmin\ radial bins, while the thin red lines marks the 16th and 84th percentiles of the [Fe/H] distribution.}
\label{r_met}
\end{figure}

 Fig.~\ref{r_met} highlights that, in the limited radial range covered by our data, we do not see significant signs of a metallicity gradient. The asymmetry and the two-component nature of the overall MDF is strikingly evident here. 

%%%%%%%%%%%%%%%%%%%%%%%%%%%%%%%%%%%%%%%%%%%%%%%%%%%%%%%%
%%%%%%%%%%%%%%%%%%%%%%%%%%%%%%%%%%%%%%%%%%%%%%%%%%%%%%%%

\begin{figure} [!ht]
 \centering
\includegraphics[width=0.8\columnwidth]{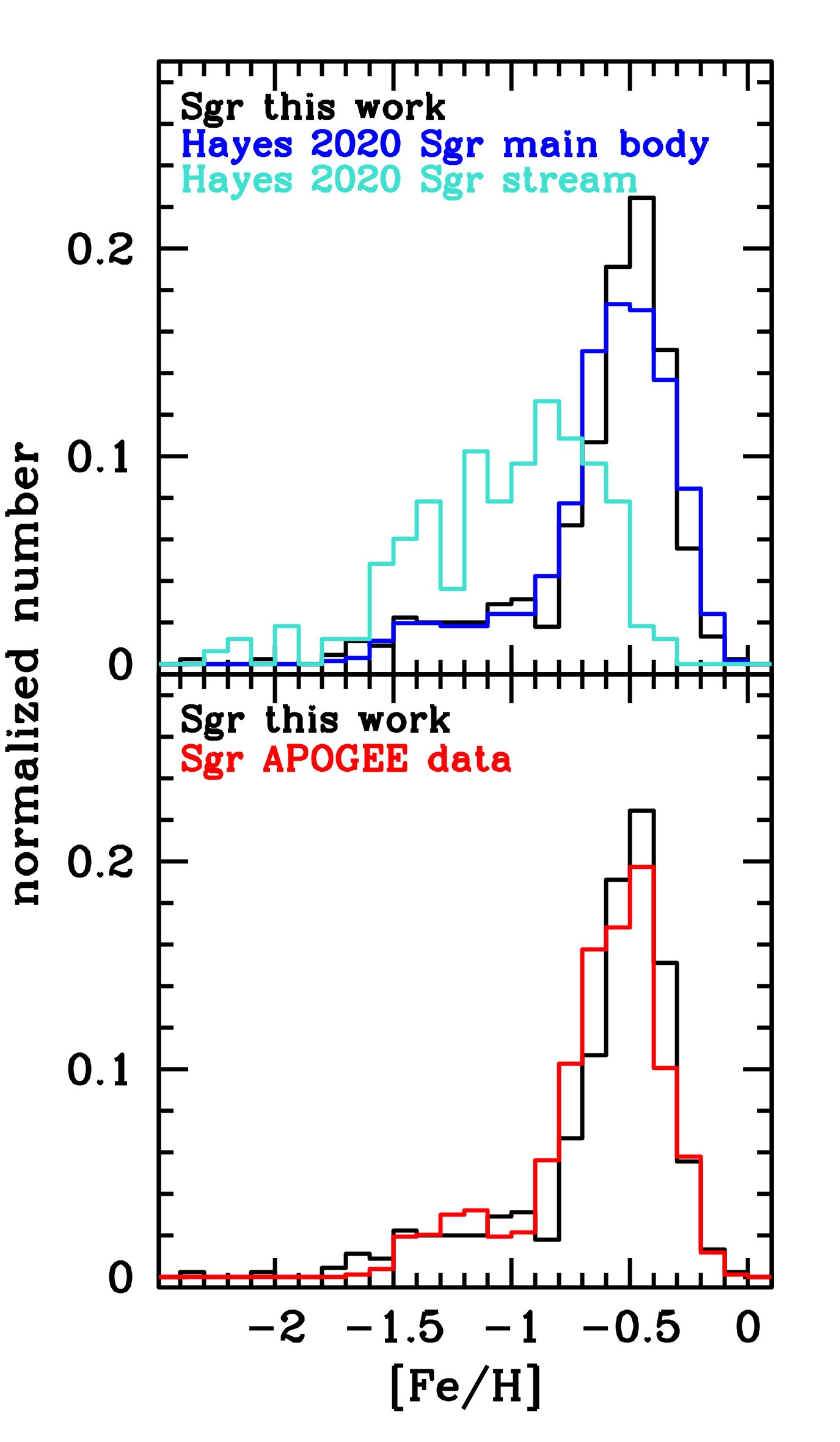}
\caption{Upper panel: comparison between the metallicity distribution derived in this work (black line) with the ones derived by \citet[][H20 samples]{Hayes2020} for Sgr main body (blue line; 710 stars) and Sgr stream stars (light blue line; 166 stars). Lower panel: the same kind of comparison but with Sgr APOGEE DR17 H21 sample (red line; 1034 stars). The distributions are normalised to have unit area.}
\label{confr_mdf}
\end{figure}

In the upper panel of Fig.~\ref{confr_mdf} we compare our MDF with those derived by \citet[][H20 hereafter]{Hayes2020} from {\sl APOGEE} \citep{Majewski2017} spectra of a large sample of Sgr stars belonging both to the main body and the stream, selected according to their angular momentum. The agreement with the H20 main body sample is quite good. The same is true for the larger sample of APOGEE DR17 \citep{APO_DR17} Sgr members that we selected following \citet[][H21 sample hereafter]{Hasselquist2021}, shown in the lower panel of Fig.~\ref{confr_mdf}. In addition to the general criteria described in Sect.~3 of \citet{Hasselquist2021}, we selected Sgr members according to their RV (100$<$RV$<$180 km/s), their distance from the Sgr centre (within 5.0\degr, and outside 15.0\arcmin, to avoid the nuclear region), their Gaia proper motions (same selection adopted for our sample), and, finally, their position on the CMD, excluding stars that were obvious outliers of the RGB.

In the main body, the overall agreement between our MDF and those shown in Fig~\ref{confr_mdf} suggests that the metallicity biases affecting the considered APOGEE samples are quite mild, and, in general, there are no large unseen populations of metal-poor stars hidden in the Sgr core \citep[as suggested also by the abundance of blue horizontal branch stars, see][and references therein]{Monaco2003}. Still, a small bias against metal-poor stars is there, and can be measured. The H20 main body sample and the H21 sample contain 710 and 1034 bona-fide Sgr member stars, respectively, but no star with [Fe/H]~$<-2.0$, while we have two such stars over 450 ($0.4\pm 0.3$\%).
The fraction of stars with [Fe/H]~$<-1.5$ ([Fe/H]~$<-1.0$) is $1.5\pm 0.5$\% ($11.5\pm 1.3$\%)
and $0.5\pm 0.2$\% ($10.8\pm 2.7$\%) in the H20 main body and H21 samples, respectively, to be compared with $2.9\pm 0.8$\% ($13.8\pm 1.9$\%) in our sample. The metal-poor population is slightly under-represented in the H20 and H21 samples with respect to ours, in spite of the much larger radial range spanned by their stars \citep[as the large scale radial gradient should make metal-poor stars more frequent at large distances from the galaxy centre,][while we are sampling the most central part of the main body core]{pristine_sgr}.

On the other hand, the comparison with the H20 MDF of the stream confirms the presence of a strong difference between the chemical composition of stars in the main body and the stream, suggesting that a large fraction of the most metal-poor component of the Sgr progenitor was stripped from the main body during the tidal disruption process \citep[][and references therein]{Chou2007,Monaco2007,LM10,Gibbons2017,Hayes2020,Johnson2020}.
In particular, the selection-corrected MDF of the Sgr stream by \citet{Johnson2020} has a mean [Fe/H]$=-0.99$, very few stars with [Fe/H]$>-0.5$ and 3\% of stars with [Fe/H]$<-2.0$. Hence, very metal-poor stars are a factor of $\simeq 7$ more abundant in their stream sample than in our main body sample.

\subsection{Comparison with other MW satellites}

\begin{figure} [!ht]
 \centering%
\includegraphics[width=\columnwidth]{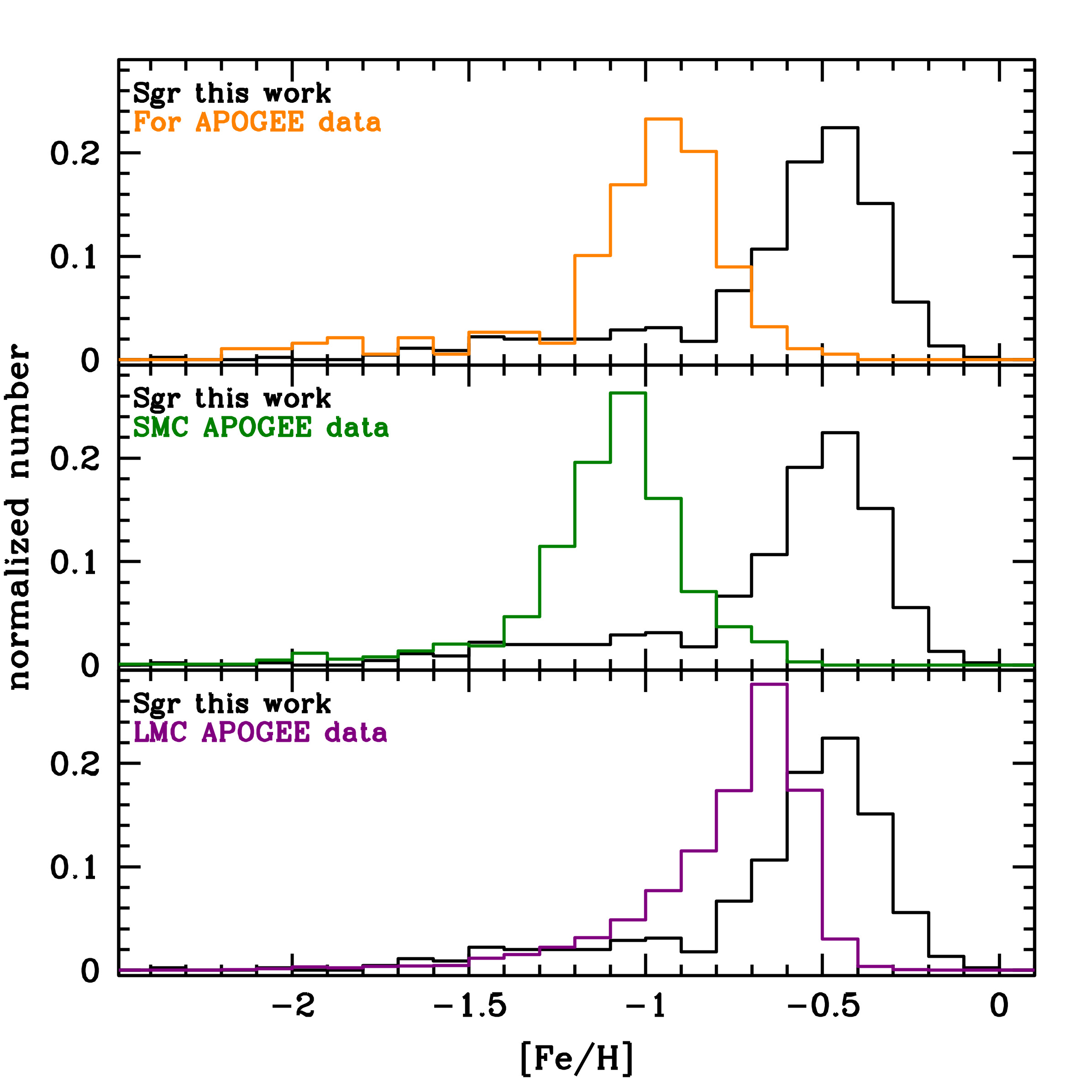}
\caption{Comparison between the metallicity distribution derived in this work (black line) with the ones from APOGEE DR17 data for different galaxies: For (orange line; 189 stars) in the upper panel, SMC (green line; 1031 stars) in the middle panel and LMC (violet line; 3897 stars) in the lower panel. The distributions are normalised to have unit area.}
\label{confr_apogee}
\end{figure}

In Fig.~\ref{confr_apogee} we compare our Sgr MDF with those of three dwarf satellites of the MW spanning a range of stellar masses \citep[taken from][]{Mcconnachie2012}, that brackets the value for the Sgr main body by \citet{vb20}:
from about 10 times lower ($2.0\times 10^7~M_{\sun}$, Fornax dSph)\footnote{Note, however, that \citet{Mcconnachie2012} reports $M_{\star}=2.1\times 10^7~M_{\sun}$ for Sgr dSph.}, to similar  ($4.6\times 10^8~M_{\sun}$, Small Magellanic Cloud, SMC hereafter) and to $\ga 10$ larger than it ($1.5\times 10^9~M_{\sun}$, Large Magellanic Cloud, LMC hereafter). The latter value is probably overestimated \citep[see][]{vb20} but here we are mainly interested to provide an idea of the involved mass ranking and range. In these cases the samples were selected from APOGEE DR17 exactly as done in \cite{Hasselquist2021}.

The MDFs of the four galaxies are fairly similar in shape: they all have a strong and relatively narrow peak on the metal-rich side, within $\la 0.5$~dex of their most iron rich stars, and a weak tail extending in the metal-poor regime. The main difference lies in the position of the peak. In this sense, the comparison is not fair, as the various samples cover hugely different radial ranges within the different galaxies and we know that the disruption process has preferentially removed metal-poor stars from the progenitor of Sgr. Still, it is remarkable that the peak of the MDF of the main body of Sgr is significantly more metal-rich than that of the other galaxies considered here, and, e.g., the fraction of stars with [Fe/H]$\ge -0.5$ is much larger than that observed in the MDF of the LMC.
This provides some support to the idea of a fairly massive progenitor of Sgr, possibly comparable to the LMC \citep[][M17]{Monaco2005,Lokas10,NiedersteOstholt2012, deBoer2014, Gibbons2017, Carlin2018, Minelli2021a, Johnson2020}.

\subsection{Comparison with chemical evolution models}
\label{chemod}

The MDF is an important constraint in chemical evolution studies. In particular, it allows to break severe degeneracies in chemical evolution model parameters that would remain in place if only the run of abundance ratios with metallicity were considered \citep[see, e.g.,][their Figs.~1 and 2]{roma15}. It is well known, in fact, that the effects of changes in several free parameters of the models tend to cancel out when the ratios of the abundances of different chemical species are considered \citep{tosi88}.

\begin{figure} [!ht]
 \centering%
\includegraphics[width=\columnwidth]{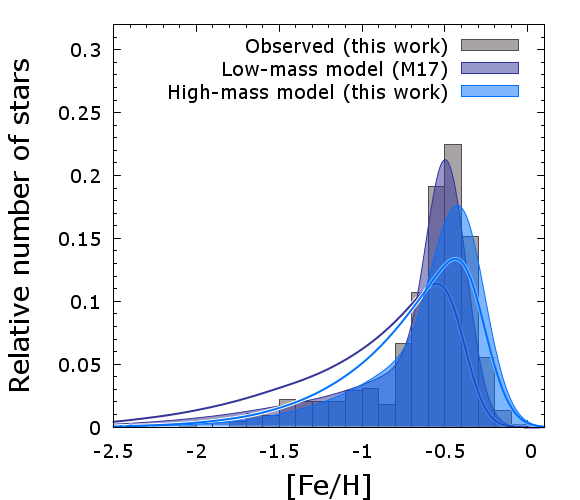}
\caption{Comparison between the MDF derived in this work (grey shaded histogram) and those predicted from the chemical evolution models by M17 and its revised version presented here. 
The MDF from the original models are presented as empty curves, while the colour shaded curves show the same distributions after removal of 75\% (M17 model) and 50\% (new model) of the metal-poor stars ([Fe/H]$<-0.6$). The distributions are normalised to have unit area.}
\label{fig:MDFGCE}
\end{figure}

Figure~\ref{fig:MDFGCE} shows the comparison between the  MDF of the core of Sgr derived in this work and those predicted by two chemical evolution models that prove able to fit the main chemical properties of Sgr. The models are the one presented and discussed in M17, and a new version of the same model slightly modified  to explore the case of a more massive progenitor. In the following, we recall their main properties \citep[see][for more details and definitions]{roma15}.

First we note that our models aim at reproducing the chemical properties of the progenitor of the Sgr system. The bulk of the stellar populations of Sgr are assumed to have formed from 14 to 7 Gyr ago (see M17 for discussion and references). The Sgr progenitor is then assumed to start losing a large fraction of its gas 
due to the interaction with the MW, until the star formation definitively stops, about 6 Gyr ago. Figure~13 of M17 shows that the run of several abundance ratios with [Fe/H] can be reasonably well explained within the above evolutionary framework. 

%%%%%%%%%%%%%%%%%%%%%%%%%%%%%%%%%%%%%%%%%%%%%%
%%%%%%%%%%%%%%%%%%%%%%%%%%%%%%%%%%%%%%%%%%%%%%

In M17, it was assumed that ~$2 \times 10^9$~M$_\odot$ of gas are accreted on short timescales, $\tau = 0.5$ Gyr, and converted into stars with an efficiency $\nu = 0.1$ Gyr$^{-1}$, till the progenitor reaches a mass in stars of $\sim 7 \times 10^8$~M$_\odot$. At this point, the satellite starts to be stripped by the Milky Way. Here, we assume a larger initial gaseous mass for the progenitor, namely, $4.2 \times 10^9$~M$_\odot$, a longer timescale for gas accretion, $\tau =$ 3 Gyr, and a higher efficiency of conversion of gas into stars, $\nu = 0.2$ Gyr$^{-1}$. The mass in stars at the beginning of the interaction with the Milky Way is $1.6 \times 10^9$~M$_\odot$, i.e., we are dealing with a LMC-like Sgr precursor. While we are still able to fit the same [X/Fe]--[Fe/H] trends as M17, we now also produce a theoretical MDF peaking at the right [Fe/H] value spotted out by the current observations, and with much less power in the metal-poor tail than the M17 model. 

In any case, both models over-predict the fraction of metal-poor stars with respect to the observed MDF, providing additional support to the hypothesis that the process of tidal stripping removed preferentially metal-poor stars from the Sgr progenitor \citep[see, e.g.,][and references therein]{LM16}. To provide a qualitative idea of the effects of this process, in Fig.~\ref{fig:MDFGCE} we also show the MDF of the two models after removal of 50\% and 75\% of the stars with [Fe/H] $< -0.6$ from the new model and the M17 model, respectively. The metal-poor side of the observed MDF can now be broadly matched and the main peak become stronger and narrower, more similar to its observed counterpart. 

The above comparison should be considered only as an insightful exercise, showing that the observed MDF is consistent with a simple but physically motivated chemical evolution model plus preferential stripping of metal-poor stars. The results would not change much if, for example, the MDF from the H21 sample would be considered instead of the one derived here, as the differences are small (Sect.~\ref{mdf}). It must be realised that a fully quantitative comparison with chemical evolution models is clearly impossible for a disrupting galaxy with a metallicity gradient. 
This formidable task could be attempted only with a self-consistent chemo-dynamical model following the evolution of all the components of the progenitor (gas, stars, dark matter), its star formation history,  chemical evolution history, the interaction of its gas component with the Milky Way disc and corona, and the process of its tidal disruption.

\subsection{Correlation between kinematics and metallicity}

In addition to a metallicity gradient within the main body and along the stream, recently \citet{Gibbons2017} found that the typical MDF in the stream is made up of two components, and that the metal-poor (MP) component  is kinematically hotter than its metal-rich (MR) counterpart.
On the other hand, \citet{Johnson2020} found that the kinematics of the most metal-poor component ([Fe/H]$<-1.9$) in their Sgr stream sample significantly differs from that of the bulk of more metal-rich component both in terms of velocity dispersion and of mean velocity in the Galactic reference frame. Chemical abundance differences between the different branches and sub-branches of the Sgr stream are discussed in detail in \citet{Ramos2021}.

The generally accepted interpretation of these features is that there was a metallicity gradient in the progenitor of Sgr, as observed in most dwarf galaxies \citep{harbeck01,taibi22}, with 
the MP component being more extended and with a 
higher velocity dispersion than the MR one, as observed, e.g., in the Sculptor dSph \citep{eline04}, Fornax dSph \citep{Batta06}, and other local dwarf galaxies \citep[see][for a recent review]{battanip22}.
 For its structural and kinematic properties, presumably, the metal-poor component was preferentially stripped from the progenitor in the early phases of its disruption, in a progressive peeling off of its composite stellar population.
 In particular, \cite{Johnson2020} suggest that the diffuse MP population originates from an extended metal-poor halo that surrounded the progenitor of Sgr.
 
 On the opposite side of the huge range of scales spanned by the Sgr system, in the innermost region of the stellar nucleus (R$\le 3.5\arcmin \simeq 25$~pc\footnote{Please note, however, that the velocity dispersion curves and rotation curves obtained by \citet{Alfaro2020} are limited to R$\le 2.0\arcmin \simeq 15$~pc, essentially for reasons of binning.}) \citet{Alfaro2020} found significant differences in the kinematics of the three components identified there \citep[see also][]{Bellazzini2008,Carlberg2022}. However, two of these components, namely the Young Metal-Rich (YMR) and the Old Metal-Poor (OMP) ones, according to their nomenclature, are characteristic of the nuclear cluster and confined within the narrow region that we purposely avoided to get an un-biased MDF representative of the main body. In particular, the OMP in the nuclear region is dominated by the globular cluster M~54 \citep{Carlberg2022}. The only component that can be related to the stellar population studied here is their Intermediate-age Metal-Rich (IMR) population, that can be considered as the innermost counterpart of our MR population, albeit the actual selection windows are not strictly equivalent. 

While these chemo-kinematic differences have been analysed in some detail in the Sgr stream \citep{Gibbons2017,Johnson2020} and in the stellar nucleus \citep{Alfaro2020}, the situation in the main body is much less explored. Indeed, the only hint to a possible difference in velocity dispersion between the MR and MP components in the main body was provided by M13. Using a sample of 328 members within $\simeq 1\degr$ from the centre of the dwarf galaxy, they highlighted a 
systematic difference in the velocity dispersion curves of the stars above and below [Fe/H]$=-0.4$, the most MP sub-sample displaying slightly higher values of the velocity dispersion than the MR one over the entire radial range covered. 

We use our new dataset to follow-up this result. With respect to M13, our sample is slightly larger and confined to a more restricted radial range. However, as expected, it provides a significantly extended sampling of the metal-poor population, as only six of the stars analysed by M13 reach [Fe/H]<-1.0 (see their Fig.~2h).

As a first step, we compare the RV distributions of the sub-samples with metallicity above or below a certain threshold [Fe/H]$_{tresh}$, using a Kolmogorov-Smirnov (KS) statistical test. The probability that the two sub-samples are extracted from the same parent RV distribution is lower than 5.0\% for thresholds in the range $-0.6\le {\rm [Fe/H]}_{tresh}\le -1.0$, with a minimum of $P_{KS}<0.5\%$ at ${\rm [Fe/H]}_{tresh}=-0.6$. We adopt this value to divide our sample into a MR sub-sample ([Fe/H]$\ge -0.6$; 294 stars) and a MP sub-sample ([Fe/H]$< -0.6$; 156 stars).  

%###################################################################
\begin{table*}
\caption{Median values of the parameters of the kinematic models}
\label{tab:mcmc}
\centering
%\hskip -1cm
\begin{tabular}{lccc}
\hline
\hline
par                       &MR sample       & MP sample        &  All   \\
$N_{*}$                   & 294            & 156              & 450   \\
$\langle RV\rangle$~[km/s]& $141.8\pm  0.6$ & $144.4\pm  1.0$ & $142.8\pm  0.5$	\\
$\sigma_{int}$~[km/s]     &  $10.5\pm  0.4$ &  $13.3\pm  0.8$ &  $11.6\pm  0.4$	\\
$V_{rot}$~[km/s]          &   $3.1\pm  0.9$ &   $1.8\pm  1.2$ &   $2.0\pm  0.8$	\\
PA[deg]                   & $116.3\pm 17.5$ &  $57.2\pm 42.5$ &  $94.7\pm 24.2$ \\
$\langle pmra\rangle$~[mas/yr]& $-2.673\pm 0.007$ & $-2.694\pm0.011  $ & $-2.680\pm 0.006$\\
$\langle pmdec\rangle$~[mas/yr]& $-1.387\pm 0.006$ & $-1.400\pm 0.009$ & $-1.392\pm 0.005$\\
$\sigma_{int}^{pmra}$~[mas/yr]& $0.103\pm 0.006$ & $0.137\pm 0.009$ & $0.117\pm 0.005$\\
$\sigma_{int}^{pmdec}$~[mas/yr]& $0.087\pm 0.005$ & $0.118\pm 0.007$ & $0.100\pm 0.004$\\
$\rho$                         & $0.255\pm 0.080$ & $0.123\pm 0.088$ & $0.195\pm 0.060$\\
\hline
\hline
\end{tabular}
\tablefoot{The median $\pm \sigma$ of the posterior PDF are reported, where $\sigma$ is estimated as half the interval between the 16-th and 84-th percentile of the distribution. All the marginalised PDFs are bell-shaped and remarkably symmetric.}
\end{table*}

%###################################################################

To gain a deeper insight on the nature of the kinematic differences between the two sub-samples we analysed them separately, as well as the entire sample, using a Bayesian approach. Monte Carlo Markov Chains (MCMC) were used to explore the posterior Probability Density Function (PDF) of the four parameters of the simple Gaussian model with rotation described by \citet{kamann18}: mean $\langle RV\rangle$, intrinsic velocity dispersion $\sigma_{int}$, amplitude of rotation $V_{rot}$ along the line of sight and the position angle of the projected rotation axis PA, measured from North toward East.
We adopted a broad Gaussian prior for $\langle RV\rangle$, with mean $\mu=143.0$~km/s and $\sigma=100.0$~km/s, uniform priors in the range 0.0-30.0~km/s for both $\sigma_{int}$ and $V_{rot}$, and a uniform prior in the range 0.0\degr-180.0\degr ~for PA. To sample the posterior PDF we used {\tt JAGS}\footnote{\tt http://mcmc-jags.sourceforge.net}, within the {\tt R} environment, to run five independent MCMCs of 20000 steps each, after a burn in phase of 1000 steps for each chain.
Results on rotation, and in particular the constraints on the position angle, must be considered with caution, due to the non uniform distribution in azimuth of our sample (see Fig.~\ref{mappa_Sgr}) that may lead to odd sampling of the underlying velocity field.
Our purpose is just to look for differences between the kinematics of the two sub-samples also in terms of ordered motions.

%###################################################################
\begin{figure*}[!ht]
\centering
\includegraphics[scale=0.31]{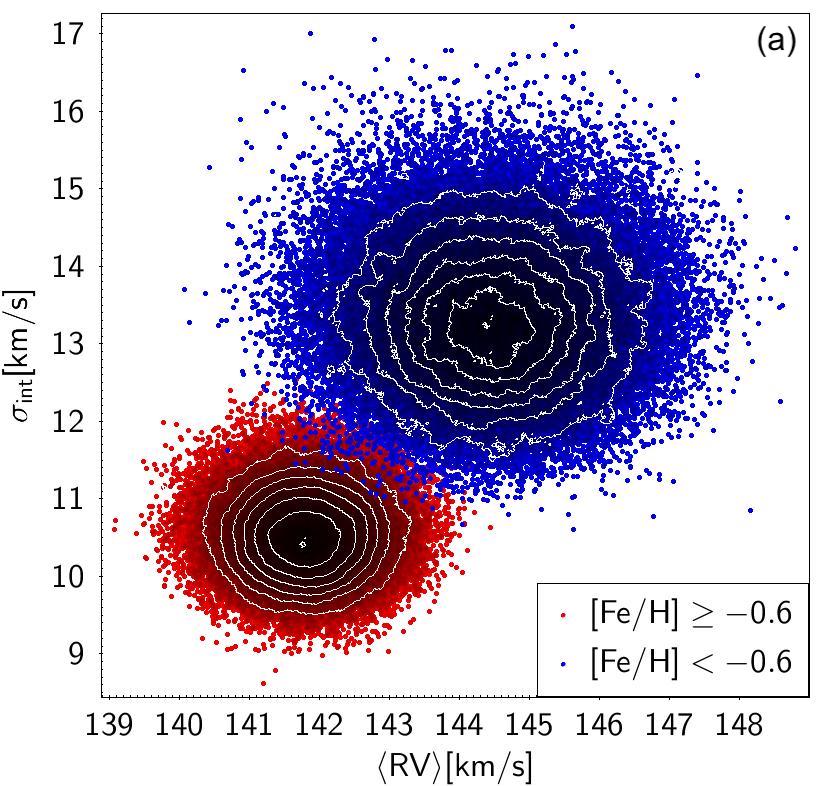}
\includegraphics[scale=0.31]{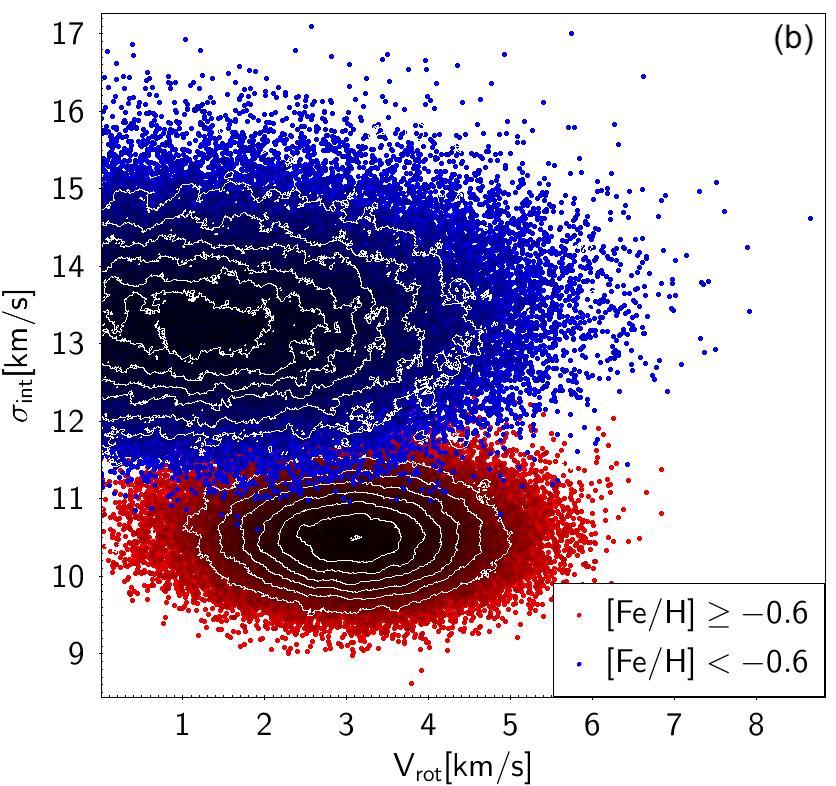}
\caption{2D posterior PDF of key parameters of the adopted kinematic model for the MP (blue points) and the MR (red points) sub-samples, as sampled with MCMC. Panel (a): $\langle RV\rangle$ and $\sigma_{int}$. Panel (b): $V_{rot}$ and $\sigma_{int}$.
\label{disp_vel}
}
\end{figure*}
%###################################################################
%%%%%%%%%%%%%%%%%%%%%%%%%%%%%%%%%%%%%%%%%%%%%%
\begin{figure} [!ht]
 \centering%
\includegraphics[width=\columnwidth]{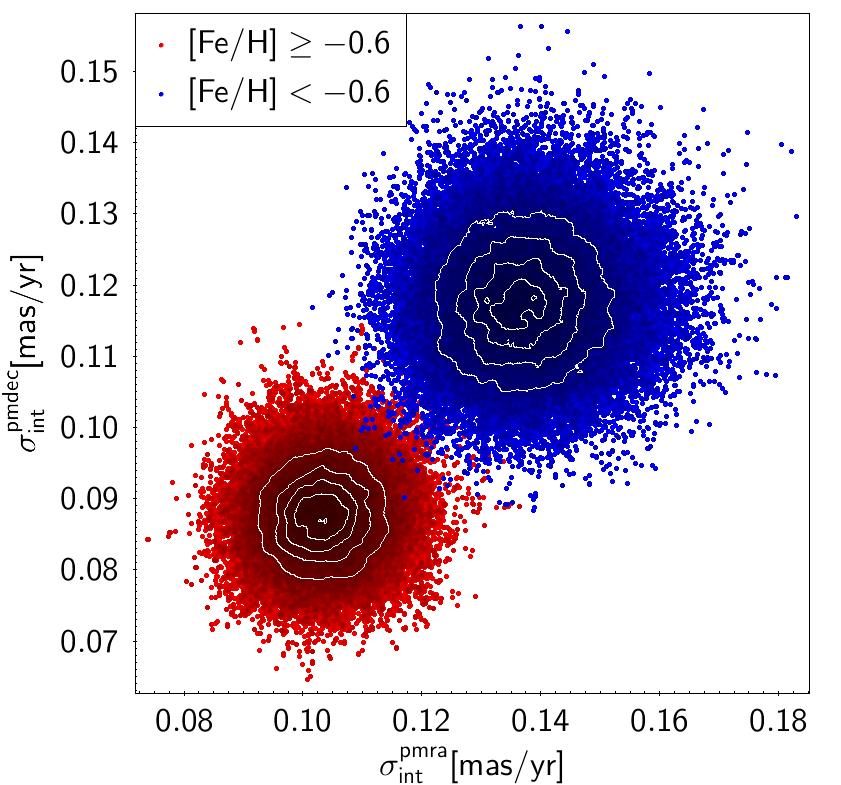}
\caption{2D posterior PDF of the proper motion dispersions for the MP (blue points) and the MR (red points) sub-samples, as sampled with MCMC.}
\label{disp_pm}
\end{figure}
%%%%%%%%%%%%%%%%%%%%%%%%%%%%%%%%%%%%%%%%%%%%%%

The main results are summarised Table~\ref{tab:mcmc} and illustrated in Fig.~\ref{disp_vel}. While the possibility that the two sub-samples are adequately described with models having the same parameters cannot be completely ruled out, their PDFs are remarkably different. In particular the velocity distribution of MP stars points to a larger velocity dispersion than the MR sample. On the other hand the observed MR velocities are hardly compatible with null values of $V_{rot}$, while the rotation signal is null or weak in the MP sample\footnote{Here we are comparing sub-samples spanning the same, small, radial range, hence perspective rotation \citep{vb20} cannot play any role in the detected difference in the PDFs of $V_{rot}$ between the MR and MP sub-samples.}. This confirms the earlier findings by M13 about the correlation between velocity dispersion and chemical composition in the Sgr core. Given the different radial range and the presence of the stellar nucleus, the comparison with the results 
of \citet{Alfaro2020} is much more difficult. 
Considering their IMR component as an approximate counterpart of the MR sub-sample considered here, it is interesting to note that they found it displays a weak rotation ($\sim 2$~km/s), in broad agreement with our results. On the other hand they find that the IMR, within $1.7\arcmin$ from the centre, has everywhere $\sigma\ga 14$~km/s, significantly higher than what found here and elsewhere at outer radii \citep[$\sigma\la 11$~km/s,][M13]{Bellazzini2008, Carlberg2022}, possibly hinting to a local deepening of the potential well in the innermost nuclear regions \citep[see also][]{Ibata2009}.

We used MCMC also to explore the proper motion distributions of the MR and MP sub-samples, taking into account the uncertainties and the correlations between the two components of the individual Gaia measures ({\tt pmra\_pmdec\_corr}). In this case the underlying model is a 2D Gaussian distribution with five parameters: $\langle pmra\rangle$, $\langle {pmdec}\rangle$, $\sigma_{int}^{pmra}$, $\sigma_{int}^{pmdec}$, and $\rho$, where $\rho$ is the correlation coefficient. Uniform priors were adopted for all the parameters, informed on the literature \citep{Helmi2018,vb20}: the means are set to range between $-5.0$~mas/yr and $0.0$~mas/yr, the intrinsic dispersions between $0.0$~mas/yr and $0.5$~mas/yr, and $\rho$ between $-1.0$ and $1.0$. The posterior PDF was sampled with 30 independent MCMCs of 6000 steps each, after a burn in phase of 1000 steps for each chain. The main results of this analysis are summarised in Table~\ref{tab:mcmc} and presented in Fig.~\ref{disp_pm}. The MP component displays significantly higher values of the velocity dispersion than the MR one in both components of the motion in the plane of the sky, thus mirroring the result form for the RV distributions. Moreover, while the correlation is compatible with zero for MP stars, it is significantly  larger than zero ($\ge 3\sigma$) in the MR sub-sample, possibly hinting to faster rotation of the MR component also in the plane of the sky.

In conclusion our analysis strongly supports the hypothesis that the chemo-dynamics differences observed in the Sgr Stream  originated in the progenitor and are still imprinted in the kinematics of the present-day main body. Our results, purely differential, limited to a small radial range and to line of sight velocity, show a non-rotating and dynamically hot MP component and a colder but weakly rotating MR component, possibly, intriguingly, hinting to a \citep[thick,][]{sanchez10} disc + halo structure of the original Sgr progenitor \citep[see also][]{Mayer2001,Kaza11,Delpino21,Carlberg2022}.
 It is interesting to note that a rotating progenitor may help to explain the observed bifurcation of the stream \citep{Pena2010,oria22,Carlberg2022}.

\section{Summary and conclusion}
\label{conclu}

We have presented the results of a spectroscopic study aimed at obtaining the metallicity distribution in the core of the Sgr dSph, outside of the nuclear region ($88$~pc$\la R\la 277$~pc).
The sample of RGB stars was carefully selected to avoid any metallicity bias. The adopted selection technique resulted in a 100\% efficiency in identifying bona-fide Sgr members, as the membership of all 450 observed stars was confirmed with radial velocity. 

The MDF we derived is characterised by a strong metal-rich peak at [Fe/H]$\simeq -0.5$ plus a wide and weak tail reaching [Fe/H]$\le -2.0$. According to a Gaussian mixture model analysis, the MD is best reproduced by two Gaussian distributions having mean and standard deviation ($\mu_1$,$\sigma_{1,int}$)=(-0.48,0.11) and ($\mu_2$,$\sigma_{2.int}$)=(-1.08,0.37), respectively, the first component accounting for $\simeq$75\% of the entire sample. Physically motivated chemical evolution models reproducing the observed abundance pattern of Sgr dSph (M17) are unable to describe adequately the observed MDF without assuming the removal of a significant fraction of metal-poor star, through the selective tidal stripping that has been invoked to explain the metallicity gradient along the Sgr stream \citep[][and references therein]{LM16,Gibbons2017,Johnson2020}.

The most recent MDFs from large APOGEE samples \citep{Hayes2020,Hasselquist2021} compares reasonably well with ours, showing that they are only mildly affected by metallicity biases. However, we showed that all stars more metal-poor than [Fe/H]$=-1.0$ are slightly under-represented in these samples, with respect to ours.

We confirm an earlier result by M13 on the presence of a correlation between the chemical abundance and the kinematic properties of the Sgr dSph stars. In particular we find that the RV distribution of stars having [Fe/H]$\ge -0.6$ clearly favours lower  values of the velocity dispersion and larger values of rotation velocity than those found for the complementary metal-poor sub-sample
([Fe/H]$< -0.6$). The MP population displays significantly higher velocity dispersion than the MR one also in each component of the motion in the plane of the sky.
This results mirror recent findings concerning samples in the Sgr stream \citep[in particular]{Gibbons2017}, suggesting that the progenitor of Sgr may have hosted a more-compact metal-rich disky component surrounded by a dynamically hotter and more extended metal-poor halo.

The results presented here provide independent assessment of and a deeper insight on the chemo-dynamical properties of the stars in the core of the Sgr dSph galaxy, a further step toward the full characterisation of this exceedingly interesting and complex system.

\begin{acknowledgements}
 The data presented and analysed in this paper were collected under the ESO program 105.20AH.001.
 
 MB acknowledge the financial support to this research by the PRIN-INAF Grant 1.05.01.85.14 {\em Building up the halo: chemo-dynamical tagging in the age 
of large surveys} (P.I. Sara Lucatello).

This work has made use of data from the European Space Agency (ESA) mission {\it Gaia} (\url{https://www.cosmos.esa.int/gaia}), processed by the {\it Gaia} Data Processing and Analysis Consortium (DPAC,\url{https://www.cosmos.esa.int/web/gaia/dpac/consortium}). Funding for the DPAC has been provided by national institutions, in particular the institutions participating in the {\it Gaia} Multilateral Agreement. 

This work has made use of Sloan Digital Sky Survey IV (SDSS-IV) data.
Funding for the SDSS-IV has been provided by the Alfred P. Sloan Foundation, the U.S. Department of Energy Office of Science, and the Participating Institutions. SDSS-IV acknowledges support and resources from the Center for High-Performance Computing at the University of Utah. The SDSS website is \url{www.sdss.org}.

SDSS-IV is managed by the Astrophysical Research Consortium for the Participating Institutions of the SDSS Collaboration including the Brazilian Participation Group, the Carnegie Institution for Science, Carnegie Mellon University, the Chilean Participation Group, the French Participation Group, Harvard-Smithsonian Center for Astrophysics, Instituto de Astrofísica de Canarias, The Johns Hopkins University, Kavli Institute for the Physics and Mathematics of the Universe (IPMU)/University of Tokyo, the Korean Participation Group, Lawrence Berkeley National Laboratory, Leibniz Institut für Astrophysik Potsdam (AIP), Max-Planck-Institut für Astronomie (MPIA Heidelberg), Max-Planck-Institut für Astrophysik (MPA Garching), Max-Planck-Institut für Extraterrestrische Physik (MPE), National Astronomical Observatories of China, New Mexico State University, New York University, University of Notre Dame, Observatário Nacional/MCTI, The Ohio State University, Pennsylvania State University, Shanghai Astronomical Observatory, United Kingdom Participation Group, Universidad Nacional Autónoma de México, University of Arizona, University of Colorado Boulder, University of Oxford, University of Portsmouth, University of Utah, University of Virginia, University of Washington, University of Wisconsin, Vanderbilt University, and Yale University.

\end{acknowledgements}

\bibliographystyle{aa}
\bibliography{AMinelli_Sgr}

\end{document}